\documentclass[prl,twocolumn,showpacs,preprintnumbers,amsmath,amssymb,superscriptaddress]{revtex4}

\usepackage{graphicx}
\usepackage{dcolumn}
\usepackage{bm}
\usepackage{mathrsfs}
\usepackage{subfigure}
\usepackage{natbib}

\begin{document}

\title{Formation of multi-quanta vortices in superconductors: \\electronic, calorimetric and magnetic evidence}

\author{Ben Xu}
\affiliation{Departement Fysica, Universiteit Antwerpen,
Groenenborgerlaan 171, B-2020 Antwerpen, Belgium}
\author{M. V. Milo\v{s}evi\'{c}}
\affiliation{Departement Fysica, Universiteit
Antwerpen, Groenenborgerlaan 171, B-2020 Antwerpen, Belgium}
\author{Shi-Hsin Lin}
\affiliation{Department of Physics, University of Notre Dame, Notre
Dame, IN 46556, USA} \affiliation{Departement Fysica, Universiteit
Antwerpen, Groenenborgerlaan 171, B-2020 Antwerpen, Belgium}
\author{F. M. Peeters}
\affiliation{Departement Fysica, Universiteit Antwerpen,
Groenenborgerlaan 171, B-2020 Antwerpen, Belgium}
\email{francois.peeters@ua.ac.be}
\author{B. Jank\'{o}}
\affiliation{Department of Physics, University of Notre Dame, Notre
Dame, IN 46556, USA}

\date{\today}

\begin{abstract}
The ground state with vorticity larger than one in mesoscopic
superconductors in applied magnetic field may manifest as a
`giant'-vortex, where all vortices coalesce into a single
singularity of the order parameter. Such a multi-quanta vortex may
split into individual vortices (and vice versa) as a function of
e.g. applied current, magnetic field or temperature. Here we show
that such transitions can be identified by heat-capacity
measurements, as the formation or splitting of a giant-vortex
results in a clear jump in measured heat capacity vs. external
drive. We attribute this phenomenon to an abrupt change in the
density of states of the quasiparticle excitations in the vortex
core(s), and further link it to a sharp change of the magnetic
susceptibility at the transition - proving that formation of a
giant-vortex can also be detected by conventional magnetometry.
\end{abstract}


\maketitle

The influence of quantum confinement on superconducting condensates
is certainly one of the prominent research directions in
low-temperature physics of the last decade. Besides the ever
intriguing properties of high-temperature superconductivity, vortex
matter in conventional but mesoscopically tailored superconducting
samples has generated tremendous interest and activity in the wide
scientific community. Indeed, vortex configurations in confined
condensates are of direct relevance to cold gases and Bose-Einstein
condensates \cite{BEC}; interaction of vortices with artificial
pinning sites has analogues in various colloidal systems and
molecular crystals \cite{molcrys}; the inhomogeneous field of
vortices may confine spin textures in a nearby magnetic
semiconductor - thus manipulation of vortex states can be useful in
spintronics \cite{natjan}; the `ratchet' dynamics of vortices in
asymmetric pinning profiles is directly related to biological
microdevices that separate particles by converting random motion
into directed motion \cite{natmem,natmos}; on the microscopic side,
the interior of a vortex core is fundamentally different from the
remainder of the superconducting sample, and may present a unique
guiding channel for applications in future electronic and optical
devices \cite{natmel}.

One of the most puzzling questions in the area of vortex matter in
submicron samples is the distinction between two allotropies of a
vortex state - a `giant' vortex, where all vortices merge into a
single singularity, and a multi-vortex, where all vortices can be
individually resolved. In type-II superconductors, transitions
between the latter two states are of second-order, following the
increasing lateral compression by e.g. increasing screening currents
in increasing magnetic field \cite{schw,geimGVS}, or increasing
temperature which makes the sample effectively smaller in terms of
the superconducting length scales. Even in numerical calculations,
it is very difficult to pinpoint the exact value of parameters for
the giant-to-multi crossover, as the order parameter is severely
suppressed between vortices in close proximity. It is therefore no
surprise that imaging experiments could not verify the existence of
a giant vortex beyond reasonable doubt \cite{irina}. Several years
ago, Kanda {\it et al.} conveyed a clever transport measurement,
where distinction between giant and multi-vortex states was made by
symmetry matching between the vortex configuration and the location
of several tunnel junctions \cite{kanda}. Although not always
conclusive \footnote{In a square with four vortices, tunnel
junctions in a symmetric arrangement would all record identical
signals regardless of the vortex state being multi or giant one.},
this is the best known method to date for giant-vortex detection.

In this Letter, we present a universal method for the observation of
formation and decay of multi-quanta vortex states. Our theoretical
simulations indicate that the experimentally {\it measured heat
capacity} of a mesoscopic superconductor as a function of magnetic
field or temperature can unambiguously reveal such transitions. The
underlying reason can be traced back to the behavior of the local
density of states for quasiparticles, and we demonstrate a direct
link between the heat capacity and the sample magnetization. With
recent advances in calorimetry \cite{prlgren} and magnetometry
\cite{geim} of submicron samples, our findings are of immediate
relevance to current experimental efforts.

The Ginzburg-Landau (GL) formalism has been extensively used in the
past to gain theoretical insight in the physics of mesoscopic
superconductors. The core of the approach is the GL energy
functional {\small
\begin{equation}
\mathcal{G}=\int\left[-|\psi|^{2}+\frac{1}{2}|\psi|^{4}+\frac{1}{2}|(-i\nabla-{\bf
A}) \psi|^{2} +\kappa^{2}({\bf h}-{\bf H})^{2} \right]dV,
\label{freeen}
\end{equation}}
describing the difference in Gibbs free energy between the
superconducting (S) and normal (N) state in units of
$\mathcal{G}_0=H_{c}^{2}\big/8\pi$. Here $\kappa$ denotes the GL
parameter and determines screening of the applied magnetic field
${\bf H}$ from the given superconducting material. In Eq.
(\ref{freeen}) all distances are scaled by the coherence length
$\xi$, the vector potential ${\bf A}$ by $c\hbar /2e\xi$, the
magnetic field ${\bf h}$ by $H_{c2}=c\hbar /2e\xi ^{2}=\kappa
\sqrt{2}H_{c}$, and the order parameter $\psi$ by its equilibrium
value in the absence of the magnetic field. The minimization of
$\mathcal{G}$ is numerically equivalent to solving two coupled GL
equations, and for details of this procedure we refer to Ref.
\cite{milgeur}. Once a stable solution is found, we are able to
calculate the specific heat of the superconducting state from the
relation $C=-T\partial^2 \mathcal{G}\big/\partial T^2$ \cite{deo},
as a difference between the total heat capacity and that of the
sample in the normal state, in units of $C_0=H^2_c(0)V/(8\pi T_c)$.
We start from the equilibrium states, increase/decrease the
temperature of the system by $10^{-4} T_c$, and calculate
numerically the second derivative. In what follows, we apply this
method for a superconducting disk, a simple geometry already
accessible both theoretically \cite{schw} and experimentally
\cite{geimGVS,irina,kanda,prlgren,geim}.
\begin{figure}[b]
\includegraphics[width=0.95\linewidth]{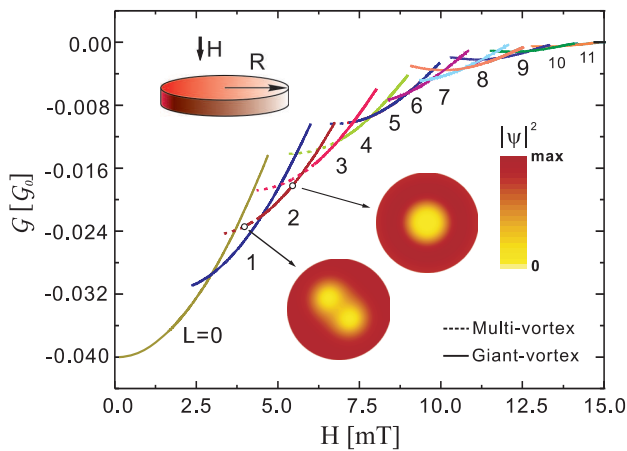}
\caption{\label{Fig1}(color online) The energy of different vortex
states for an $Al$ superconducting disk of radius $R=850$ nm and
thickness $d=100$ nm, at $T=0.8T_c$, for taken $\xi(0)=100$ nm and
$\kappa=1.2$. Solid lines indicate giant-vortex states and dashed
lines represent multi-vortex states. Insets show the density of the
superconducting condensate for a $L=2$ vortex state in multi and
giant form.}
\end{figure}

Figure \ref{Fig1} shows the energy of all the vortex states found in
an Aluminum superconducting disk of radius $R=850$ nm and thickness
$d=100$ nm, at $T=1.1$ K (we use $\xi(0)=100$ nm, $\kappa=1.2$, and
$T_c=1.38$ K \cite{golub}). The confinement effects are more
significant for increased vorticity in increasing field, due to the
interaction of the flux quanta with lateral boundaries. For that
reason, all states in Fig. \ref{Fig1} with $L>5$ are giant vortices.
However, for $2<L\leq 5$ multivortex states can be found at lower
magnetic field, which are compressed into a giant-vortex at higher
applied field. This is a gradual, second-order transition, and is
therefore invisible in the free energy curves \cite{schw}. For
clarity, we made distinction between multi- and giant-vortex in Fig.
\ref{Fig1} by dashed and solid lines respectively. In what follows,
we discuss the repercussions of the latter transition on the heat
capacity of the sample.

\begin{figure}[b]
\includegraphics[width=\linewidth]{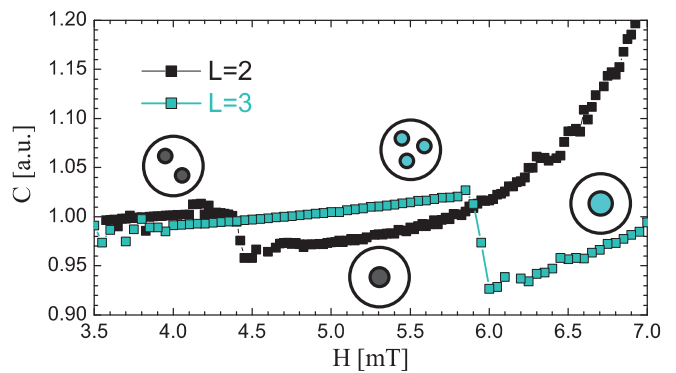}
\caption{\label{Fig2}(color online) The heat capacity as a function
of magnetic field, for states with vorticity 2 and 3 of the sample
considered in Fig. \ref{Fig1}. Insets depict the vortex
configuration before and after the multi-to-giant vortex
transition.}
\end{figure}

Using attoJoule calorimetry, Ong {\it et al.} \cite{ong} studied the
heat capacity of mesoscopic disks as a function of the magnetic
field and found that the heat capacity is directly linked to the
vorticity, exhibiting jumps at transitions between vortex states. We
argue here that the heat capacity depends not only on the number of
vortices in the sample, but also on their configuration. Namely, the
susceptibility of the sample to heating is linked to the kinetic
energy of the Cooper-pairs in and around the vortex core(s), and the
changes in their trajectory upon the multi-to-giant vortex
transition. Using the definition of heat capacity as a second
derivative of the Gibbs free energy with temperature, we calculate
it as a function of applied magnetic field for vortex states with
vorticity 2 and 3, both exhibiting multi-to-giant vortex transition
in Fig. \ref{Fig1}. As shown in Fig. \ref{Fig2} for both cases, the
general trend of increasing heat capacity with field is interrupted
{\it exactly} at the multi-to-giant vortex transition, where a sharp
decrease of heat capacity is found.

In what follows, we show that the cause of the observed change in
heat capacity during the merging of vortices is the changing local
density of states (LDOS) for quasiparticle excitations inside the
vortex cores. Actually, already from the early theoretical works
(see Ref. \cite{car}), we know that the bound state spectrum inside
the vortex is also a function of momentum along the vortex line; as
a result, the lowest bound state energy for winding number $L>1$ is
$L$ times larger than that of winding number 1. Therefore, one
expects that the low-energy states are pushed toward higher energies
during merging of individual vortices into a giant vortex. To give a
quantitative measure of this process, we first obtain the order
parameter $\psi$ and vector potential ${\bf A}$ of the equilibrium
states from the GL calculation, which then serve as inputs for the
microscopic Eilenberger equation \cite{eilen}
\begin{equation}
-i\hbar{\bf v}_F\cdot \nabla \hat{g}({\bf
r},i\tilde{\varepsilon}_n)=\left[\left[
                            \begin{array}{cc}
                              i\tilde{\varepsilon}_n & -\Delta({\bf r}) \\
                              \Delta^\dag({\bf r}) & -i\tilde{\varepsilon}_n \\
                            \end{array}
                          \right], \hat{g}({\bf
r},i\tilde{\varepsilon}_n) \right], \label{eil}
\end{equation}
where $i\tilde{\varepsilon}_n({\bf r})=i\varepsilon_n({\bf r})+{\bf
v}_F\cdot \frac{e}{c}{\bf A}(\bf r)$, and $\hat{g}=\left(
                            \begin{array}{cc}
                              ig & f \\
                              -f^\dag & -ig \\
                            \end{array}
                          \right)$ with normalization $\hat{g}({\bf
r},i\tilde{\varepsilon}_n)\hat{g}({\bf
r},i\tilde{\varepsilon}_n)=-\pi^2\hat{l}$. Eq. (\ref{eil}) is
further parameterized by $f=\frac{2a}{1+ab}$,
$f^{\dagger}=\frac{2b}{1+ab}$, and $g=\frac{1-ab}{1+ab}$, where
functions $a$ and $b$ now satisfy the independent nonlinear Ricatti
equations \cite{maki}. In the next step, the LDOS is evaluated from
\begin{equation}
 N(E,\textbf{r})= N_0 \int^{2\pi}_{0}\frac{d\theta}{2\pi}\rho(\theta) Re~\textbf{g}(i\varepsilon_n\rightarrow E+i\eta,
 \textbf{r},\theta),
\label{dos}
\end{equation}
where $\eta(>0)$ is a small real constant. To obtain
$\textbf{g}(i\varepsilon_n\rightarrow E+i\eta, \textbf{r},\theta)$,
we solve the Eilenberger equations for $\eta-iE$ instead of the
Matsubara frequency $\omega_n$. In order to find the LDOS, the above
equations should be solved for a bundle of trajectories with
different angle $\theta$, running through the given point
$\textbf{r}$ and energy $\varepsilon$. In our calculation we
consider only specular reflection for trajectories encountering the
outer boundary of the sample.

\begin{figure}[t]
\includegraphics[width=\linewidth]{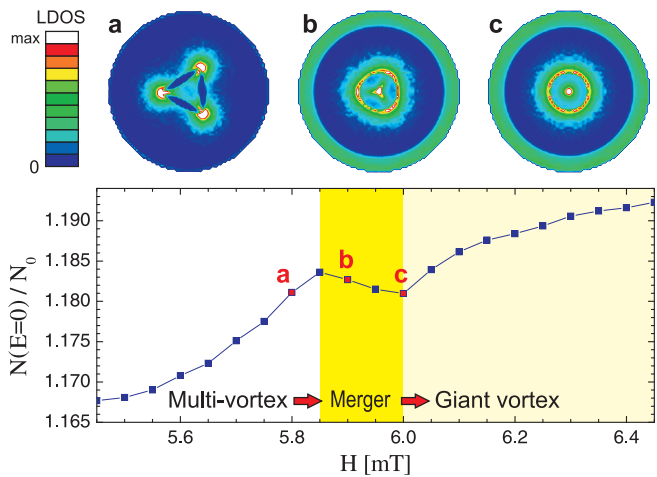}
\caption{\label{Fig3}(color online) The integrated zero-energy
density of states (LDOS) as a function of the magnetic field for
$L=3$, for a superconducting disk with same parameters as in Fig.
\ref{Fig1}. (a-c) are the representative contourplots of LDOS in the
disk, at indicated magnetic fields. Interestingly, $N(0)[a]\approx
N(0)[b]\approx N(0)[c]$.}
\end{figure}

Early theoretical works already considered LDOS of a single vortex
\cite{NNakai}. Further, in Ref. \cite{natmel} LDOS was calculated
semiclassically for a multivortex vs. the case of a giant vortex,
for selected vorticities and assumed size and distribution of the
vortex cores. Here we present the full evolution of the LDOS of
quasiparticle excitations in a mesoscopic superconducting disk,
during the multi-to-giant vortex transition as a function of the
magnetic field, where at each step we calculate the distribution of
the superconducting order parameter. In Fig. \ref{Fig3}, we plot the
zero-energy density of states $N(E=0,T)$ integrated over the sample
as a function of the applied magnetic field, for the $L=3$ vortex
configuration. $N(E=0,T)$ increases with applied magnetic field, as
in the case of an isolated vortex \cite{NNakai}. When the giant
vortex is assembled from the multi-vortex molecule, the LDOS profile
changes from several individual peaks located at each vortex to a
ring-like bound state with/without an enclosed peak for odd/even
vorticity (as in Ref. \cite{natmel}). Bound states are also found
near the sample boundary, due to there lowered (non-zero) gap in the
presence of strong circular Meissner currents. The representative
contour plots of LDOS for $L=3$ are shown in Fig. \ref{Fig3} as
insets. As our main observation, we point out a clear drop of
$N(E=0)$ vs. the magnetic field at the multi-to-giant vortex
transition (see Fig. \ref{Fig3}), where the LDOS profile goes
through a change of symmetry from three-fold to circular symmetric
one. We thus confirm that the evolution of LDOS with magnetic field
is directly linked to the specific heat and the thermal conductivity
of the sample. To enforce this argument, the LDOS can be expressed
as $N(E,T)/N_0= N(E=0,T) + \alpha_E |E|/\Delta_0$, and $N(E=0,T)/N_0
= \gamma' + \alpha_k T/T_c$. Through the relation
$C(T)/T=(2/T)\int_0^{\infty}dE [ E~N(E,T)\partial f(E,T) / \partial
T]$, using the Fermi distribution function $f(E,T)$, one obtains
$C(T)/(\gamma_n T) \sim N(E=0,T)/N_0 + \alpha_E |E|/\Delta_0 $ which
unambiguously shows the link between the calculated curves in Figs.
\ref{Fig2} and \ref{Fig3}.

\begin{figure}[t]
\includegraphics[width=0.9\linewidth]{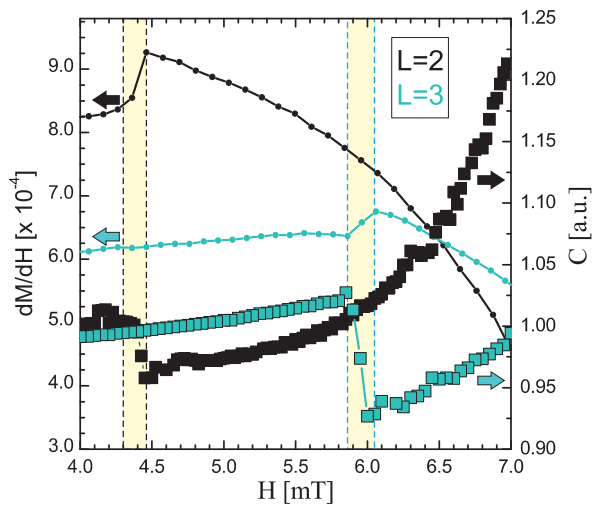}
\caption{\label{Fig4}(color online) The multi-to-giant vortex
transition revealed through the sharp change in magnetic
susceptibility as a function of applied magnetic field, showing
direct correlations with the heat capacity, for states with
vorticity 2 and 3. Shaded areas indicate the observed regions of
giant-vortex formation in both quantities.}
\end{figure}
The jump of heat capacity between different vortex phases can also
be expressed using other thermodynamic arguments. The discontinuity
in the specific heat at a phase transition (at field $H^*$) can be
calculated as \cite{pgdg}
\begin{equation}\label{eq:C_B}
C_i - C_j = -T\bigg( \frac{dH^*}{dT}\bigg)^2 \bigg[ \bigg(
\frac{\partial M}{\partial H}\bigg)_i - \bigg( \frac{\partial
M}{\partial H}\bigg)_j\bigg],
\end{equation}
where $M$ denotes sample magnetization. Here we apply above
expression to the multi-to-giant vortex transition, where $i$
represents the vortex state of vorticity $L$ just prior, and $j$
represents the $L$ vortex state just after the transition. Knowing
the result for heat capacity vs. $H$ at the multi-to-giant vortex
transition, we therefore expect to see similar features in the
magnetic susceptibility $\chi=\partial M/\partial H$. We calculate
the magnetization $M$ as expelled magnetic field from the sample
${\bf M} = (\langle {\bf h} \rangle -{\bf H})\big/4\pi$, where
$\langle {\bf h} \rangle$ is the local magnetic field averaged over
the sample volume. The results of this calculation are shown in Fig.
\ref{Fig4}. They (i) confirm the link between (independently
calculated) sharp changes in heat capacity and $\chi$ as a function
of the magnetic field, and (ii) show that assembly of a giant vortex
in superconductors can be detected even by conventional
magnetometry.

\begin{figure}[b]
\includegraphics[width=0.9\linewidth]{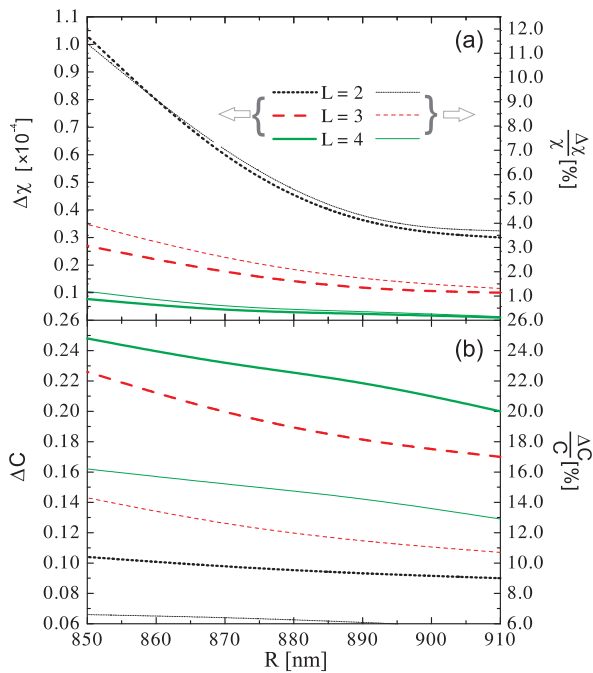}
\caption{\label{Fig5}(color online) The absolute and relative size
of the jump at the multi-to-giant vortex transition, in (a)
$\chi(H)$ curves, and (b) $C(H)$ curves, for vorticity $L=$2,3, and
4.}
\end{figure}

The above prediction is of immediate relevance to experiments, since
both calorimetry and magnetometry are readily performed on
mesoscopic superconductors. Of course, the question of sensitivity
and resolution of the measurement is an open one, and we address
this issue in Fig. \ref{Fig5}. First, we determined the
multi-to-giant vortex transition field $H^*$ as a function of the
size of the $Al$ disk. For all considered vorticities ($L=2-4$),
$H^*$ was found to increase with the radius of the sample. We then
scanned the heat capacity and magnetic susceptibility versus applied
field for every size of the sample, and recorded the size of the
observed jump between values prior and after $H^*$. In Fig.
\ref{Fig5} we show the absolute and relative size of the jump of
both magnetic susceptibility (b) and heat capacity (c) at
temperature 1.1 K. We found that the susceptibility shows a clearer
signal at the multi-to-giant transition for lower vorticity, whereas
corresponding discontinuity of heat capacity is more pronounced at
higher vorticity. Note however that $\Delta C$ and $\Delta \chi$
should be directly proportional, according to Eq. (\ref{eq:C_B}).
They indeed are, when susceptibility is calculated by
$\chi=\partial^2\mathcal{G}\big/\partial H^2$, while we here used
the experimental definition of magnetization as the flux expelled
from the sample (which is directly measured by Hall magnetometry).

In summary, we demonstrated that second-order transitions between
multi- and giant-vortex states in mesoscopic superconductors can be
detected using calorimetry. The local density of states for
quasiparticles in and around vortex cores changes when the vortex
configuration changes, which affects the heating properties of the
system. The observed sharp change in the heat capacity at the
multi-to-giant vortex transition can also be linked to the magnetic
susceptibility, enabling the observation of this transition by Hall
magnetometry. Our results are therefore of immediate relevance to
experimental efforts in the field, and further work is needed to
generalize our findings to other systems, such as e.g. Bose-Einstein
condensates \cite{BEC}.

We thank O. Bourgeois, T. Yokoyama, M. Eschrig and M. Ichioka for
discussions. This work was supported by the Flemish Science
Foundation (FWO-Vl), the Belgian Science Policy (IAP), the bilateral
project Flanders-USA, NSF NIRT, ECS-0609249, and the Institute for
Theoretical Sciences, Notre Dame.

\end{document}